\documentclass{article}
\usepackage{fleqn,ams}
\usepackage{amssymb}
\usepackage{verbatim}
\newcommand{\const}{\mbox {Const}}

\newcommand{\depode}[2]{\frac{\partial #1}{\partial #2}}
\newcommand{\imp}[1]{{\cal P}_{#1}}
\newcommand{\dvu}[1]{\partial_{#1}}
\newcommand{\Ham}{{\cal H}}

\newcommand{\Per}[1]{#1_{\perp}}
\newcommand{\Par}[1]{#1_{\parallel}}
\newcommand{\mupar}{\Par{\mu}}
\newcommand{\muper}{\Per{\mu}}
\newcommand{\mcmupar}{m^2 + \mupar^2 p^2}
\newcommand{\mcmuper}{m^2 + \muper^2 p^2}
\newcommand{\muminmu}{\mupar^2 - \muper^2}
\newcommand{\mupln}{\ln \left( \frac{\sqrt{\mcmupar} +
p \sqrt{\muminmu}}{\sqrt{\mcmuper}} \right)}
\def\stackunder#1#2{\mathrel{\mathop{#2}\limits_{#1}}}

\begin{document}

\begin{center}
{\bf\Large Kinetic model of GMSW in an anisotropic plasma} \\[12pt]
Yu.G. Ignat'ev \\
Kazan State Pedagogical University\\ 1 Mezhlauk Str., Kazan
420021, Russia
\end{center}

\begin{abstract}
A kinetic model of gravimagnetic shock waves (GMSW) in a locally
anisotropic plasma is investigated. The equations of a drift
approximation are written, and the moments of the distribution
function are calculated. Solutions of the drift equations for a
highly anisotropic ultrarelativistic plasma are found. It is shown
that in this case the GMSW essentially affect the angular
characteristics and the intensity of the magneto -
brem\-s\-strah\-lung of the magnetoactive plasma.
\end{abstract}

\section{Introduction}
In Ref. \cite{gmsw}, from the requirement that the dynamical
velocity of plasma be equal to that of the electromagnetic field
due to the Einstein equations and the first group of the Maxwell
equations, the equations of the relativistic magnetic
hydrodynamics (RMHD) of magnetoactive plasma in a gravitational
field were derived. In the same work (see also \cite{mistake}) a
remarkable class of exact solutions of the RMHD equations was
obtained; it describes the motion of a magnetoactive locally
isotropic plasma in the field of a plane gravitational wave (PGW)
and is called gravimagnetic shock waves (GMSW). In Ref.
\cite{gmsw2} it was shown that in pulsar magnetospheres the GMSW
are a highly effective detector of the gravitaional radiation of
neutron stars. An observed consequence of the energy
transformation from the gravitaional wave to the GMSW energy are
the so-called giant pulses, sporadically arising in the radiation
of a number of pulsars. The estimates made in \cite{gmsw2} and
\cite{detector} allow one to identify the giant pulses in the
pulsar NP 0532 radiation with the gravitational radiation of this
pulsar in the basic quadrupole mode of a neutron star.

A fundamental importance of GMSW for the gravitational theory
leads to the necessity of a more detailed and comprehensive
investigaton of this phenomena. As was shown in \cite{gmsw} and
\cite{gmsw2}, a GMSW is realized in an almost collisionless and
nonequilibrium plasma situated in an abnormally strong
mag\-ne\-tic field. Under these conditions, as a consequence of
strong magneto - brem\-s\-strah\-lung, the isotropy of the local
distribution of plasma electrons is significantly violated, as was
assumed in obtaining the solution in \cite{gmsw}. In \cite{hydro},
on the basis of the general RMHD equations, a hydrodynamic model
of GMSW in an anisotropic plasma was built. Since in the
hydrodynamical approach the number of equations obtai\-ned is
smaller than that of unknown functions, we had to postulate a
relation between the longitudinal and transverse components of the
plasma pressure. In \cite{hydro} the simpliest variant of such a
(linear) relation was studied. The study revealed a high
dependence of the GMSW process on the degree of plasma anisotropy,
leading to the necessity of building a dynamical model of
anisotropic magnetoactive plasma motion in the gravitaional
radiation field. It is the problem dealt with in the present
paper. Throughout the paper a set of units is used where $(c = G =
\hbar = 1)$.

\section{Drift solution to the kinetic
equation for an anisotropic plasma}
\subsection{Field quantities in the plane gravitaional
wave metric}
Let us study a collisionless plasma in the PGW metric \cite{torn}:
\begin{equation}  \label{1.1}
d s^{2}=2dudv-L^{2}[e^{2
\beta}(dx^{2})^{2}+e^{-2\beta}(dx^{3})^{2}] \equiv 2 du dv - A
(dx^2)^2 - B (dx^3)^2\,,
\end{equation}
where $\beta(u)$ is an arbitrary function (the PGW amplitude), the
function $L(u)$ (the background factor of the PGW) obeys a
second-order ordinary differential equation;
$u=\frac{1}{\sqrt{2}}(t - x^{1})$ is the retarded time, $v=
\frac{1}{\sqrt{2}}(t + x^{1})$ is the advanced time. The absolute
future corresponds to the range ${\it T^{+}}:\lbrace
u>0;v>0\rbrace $, the absolute past to ${\it T^{-}}:\lbrace
u<0;v<0\rbrace $. The metric (\ref{1.1}) admits a group of motions
${\cal G}_{5}$, with the corresponding three linearly independent
Killing vectors at a point:
\begin{equation}
\label{1.2} \stackunder{(1)}{\xi^{i}} =\delta^{i}_{v}; \qquad
\stackunder{(2)}{\xi^{i}} = \delta^{i}_{2}; \qquad
\stackunder{(3)}{\xi^{i}} = \delta^{i}_{3}\,.
\end{equation}
Let there be no GW for $u \leq 0$, i.e., --
\begin{equation}
\label{1.3} \beta(u)_{\mid u \leq 0}=0; \qquad L(u)_{\mid u \leq
0}=1,
\end{equation}
and the homogeneous magnetic field be directed in the plane
$\lbrace x^{1},x^{2} \rbrace $:
\begin{displaymath}
H_{1 \mid u \leq 0}=H_{0} \cos\Omega ; \qquad H_{2 \mid u \leq
0}=H_{0} \sin\Omega ;
\end{displaymath}
\begin{equation}
\label{1.4} H_{3 \mid u \leq 0}=0; \qquad E_{i \mid u \leq 0}=0,
\end{equation}
where $\Omega$ is an angle between the $0x^{1}$ axis (PGW
propagation direction) and the magnetic field direction ${\bf H}$.
The conditions (\ref{1.4}) correspond to the vector potential
\begin{equation}
\label{1.5} A_{v}=A_{u}=A_{2}=0;\quad A_{3}=H_{0} (x^{1}
\sin\Omega - x^{2} \cos\Omega); \quad (u \leq 0).
\end{equation}
The electromagnetic field in the PGW metric (\ref{1.1}) for an
initially homogeneous plasma is described by the vector potential
\cite{gmsw}:
\begin{displaymath}
A_2 = A_v = A_u = 0\,;
\end{displaymath}
\begin{equation}  \label{1.6}
A_{3} = - H_{0} x^{2} \cos\Omega + \frac{1}{\sqrt{2}} H_{0} [v -
\psi(u)] \sin\Omega,
\end{equation}
where $\psi(u)$ is an arbitrary differentiable function satisfying
the initial condition:
\begin{equation}
\label{1.7} \psi_{|u\leq 0} = u.
\end{equation}
In this case the only nonzero component of the Maxwell tensor
depending on $\psi$ is
\begin{equation}
\label{1.8} F_{u3} = - \frac{1}{\sqrt{2}}H_{0}\psi'\sin\Omega.
\end{equation}
Other nontrivial components of the Maxwell tensor are:
\begin{equation}  \label{1.9}
F_{23} =  - H_0 \cos \Omega \,; \qquad F_{v3} = \frac{1}{\sqrt{2}}
H_0 \sin \Omega \,.
\end{equation}

\subsection{Collisionless kinetic equation}
The collisionless kinetic equation for the 8-dimensional
distribution function for charged particles of a kind $a$,
$F_a(x^i, \imp{i})$ has the form \cite{kin}:
\begin{equation}  \label{1.10}
[\Ham_a , F_a] \equiv \depode{F_a}{x^i} \depode{\Ham_a}{\imp{i}} +
\depode{F_a}{\imp{i}} \depode{\Ham_a}{x^i} = 0\,,
\end{equation}
where
\begin{equation}  \label{1.10a}
\imp{i} = p_i + e_a A_i
\end{equation}
is the generalized momentum of a particle and
\begin{equation}  \label{1.10b}
\Ham_a(x^i, \imp{i}) = \frac{1}{2} g^{ij} \left(\imp{i} - e_a A_i
\right) \left(\imp{j} - e_a A_j \right)
\end{equation}
is the Hamiltonian function of a charged particle.

The process of obtaining equations in the drift approximation on
the basis of the collisionless kinetic equations (\ref{1.10}) was
described by the author in \cite{probl}. Here we only somewhat
transform the process for the case of an initially anisotropic
distribution. Note, besides, that in the case $\Omega \not= \pi/2$
the results \cite{probl} are erroneous because they use $\imp{2}$
in the role of an integral of motion, which is the case only if
$\Omega = \pi/2$. That was mentioned in the work cited above
\cite{gmsw}.

Let us study the case when the GW propargates perpendicularly to
the magnetic field, $\Omega = \pi /2$, and let us look for
solutions of the kinetic equation independent of the variables $v,
x^2$ and $x^3$ . As in \cite{probl}, let us introduce the unit
timelike vector $v^i$:
\begin{equation}  \label{1.11}
v_2 = v_3 = 0\,;\quad v_u =  \sqrt{\frac{\psi'}{2}}\,; \quad v_v =
\sqrt{\frac{1}{2 \psi'}}
\end{equation}
and transform the kinetic equation into the frame of reference
(FR) moving with the velocity $v^i$:
\begin{equation} \label{1.12}
\imp{v} = v_u (\imp{4} + \imp{1})\; \quad \imp{u} = v_v (\imp{4} -
\imp{1})\,.
\end{equation}
The Jacobian of this transformation is equal to unity:
\begin{equation} \label{1.13} \displaystyle
\frac{D(\imp{v}, \imp{u})}{D(\imp{1}, \imp{4})} = 1\,.
\end{equation}
As a result, we get the equation:
\begin{displaymath}
v_v (\imp{4} + \imp{1}) \depode{F_a}{u} -v'_v (\imp{4} + \imp{1})
\left(\imp{1}\depode{F_a}{\imp{4}} + \imp{4}
\depode{F_a}{\imp{1}}\right) - \frac{e H}{\sqrt{B}} \imp{3}
\depode{F_a}{\imp{1}}
\end{displaymath}
\begin{equation}  \label{1.14}
  +\frac{1}{2} \left[ (A^{-1})'\imp{2}^2 + (B^{-1})'\imp{3}^2\right]
v_u \left(\depode{F_a}{\imp{4}} + \depode{F_a}{\imp{1}}\right) =
0\,,
\end{equation}
where:
\begin{equation}  \label{1.15}
H^2 = - \frac{2}{B} \dvu{u}A_3 \dvu{v}A_3 = H^2_0 \psi'
\frac{e^{2\beta}}{L^2}
\end{equation}
is an invariant of the electromagnetic field (the magnetic field
intensity squared in the FR moving with the velocity $v^i$).

On the PGW front the solution (\ref{1.14}) must satisfy the
initial condition cor\-res\-pon\-ding to a homogeneous anisotropic
currentless plasma:
\begin{equation}  \label{1.14a}
F_a(x^i, \imp{i})_{|u = 0} = f_a (\Per{p}^2, \Par{p}^2)
\delta(\Ham_a - \frac{1}{2} m_a^2 )\,,
\end{equation}
where $f_a$ is an arbitrary function of its arguments and the
following notations are introduced:
\begin{equation}  \label{1.14b}
\Per{p}^2 = p_1^2 + p_3^2 \,; \quad \Par{p} = - p_2 \,.
\end{equation}

The kinetic equation (\ref{1.14}) has three exact integrals:
\begin{displaymath}
\Ham_a(x^i, \imp{i}) = \frac{1}{2} m_a^2  = \const \,;
\end{displaymath}
\begin{equation}  \label{1.16}
\imp{2} = \const \,; \qquad \imp{3} = \const \,.
\end{equation}
For a complete solution of the problem we need one more
independent integral.

\subsection{Drift approximation}
We shall solve Eq. (\ref{1.14}) in the drift approximation, when
the Larmor frequency for each kind of charged particles --
\begin{equation}  \label{1.17}
\omega_a = \frac{e_a H}{m_a}
\end{equation}
is much greater than the characteristic frequency $\omega$ of the
gravitational wave:
\begin{equation}  \label{1.18}
\Lambda = \frac{\omega}{\omega_a} \ll 1 \,.
\end{equation}
In the zeroth order with respect to the parameter $\Lambda$
(\ref{1.14}) takes the form:
\begin{equation}  \label{1.19}
H \depode{F_a}{\imp{1}} = 0\,,
\end{equation}
i.e. in the drift approximation $F_a$ is independent obviously of
$\imp{1}$. Thus in the drift approximation, apart from the above
exact integrals, the kinetic equation has also drift (approximate)
integrals \cite{probl}:
\begin{equation}  \label{1.20}
\imp{4} \approx \const \,; \quad u \approx \const \,.
\end{equation}
The solution of the kinetic equation corresponding to the drift
approximation, which, in the absence of GW, is transformed to an
anisotropic and currentless one, can be written in the form
\begin{equation}  \label{1.21}
F_a = f_a\left( \imp{4}^2 - \imp{2}^2 , \imp{2}^2 , u \right)
\delta(\Ham_a - \frac{1}{2} m_a^2 ) \,,
\end{equation}
where $f_a$ is an arbitrary function of its arguments. In
particular, the following fa can be chosen:
\begin{equation}  \label{1.22}
f_a = \stackrel{0}{f}_a \left[ \muper^{-2} (\imp{4}^2 - \imp{2}^2
- m_a^2) + \mupar^{-2} \imp{2}^2 \right] \,,
\end{equation}
where $\mupar(u)$ and $\muper(u)$ are arbitrary functions of their
arguments. Thus in the FR (\ref{1.12}) moving with the velocity vi
the drift solution locally coincides with the unperturbed
distribution (\ref{1.14a}).

Substituting the obtained distribution function of the zero drift
approximation (\ref{1.21}) into the kinetic equation (\ref{1.14}),
it is easy to derive a correction of the first drift
approximation, $\delta F_a \sim \Lambda^{-1}$, for the
distribution function determined up to an additive component being
an arbitrary function of the above drift integrals (see
\cite{probl}). However, exact consequences of the collisionless
kinetic equation (\ref{1.14}) are conservation laws for the number
of each kind of particles and the total energy-momentum tensor
(EMT) of the plasma and the electromagnetic field \cite{kin2}.
These laws impose certain restrictions on the above additive
component and lead to differential equations for the functions
$\mupar(u)$ and $\muper(u)$.

Due to the symmetry properties, it turns out that the first order
correction to the distribution function contributes only to the
component $n_3$ of the particle number current density vector and
the components $T^a_{v3}$ and $T^a_{u3}$ of the particle EMT.
However, the component n3(u, v) does not affect the continuity
equation, only the above two components of the particle EMT appear
in the transport equation of the total EMT for the component $i =
3$ they do not appear in other transport equations. On the other
hand, the drift current determined by the first correction to the
distribution function can be obtained as a consequence of the
energy-momentum conservation laws and the Maxwell equations,
without addressing to a solution of the kinetic equation (see
\cite{gmsw}). As a result, it turns out that the set of the
transport equations and the Maxwell equations split into two
subsets, one of which, determined by the zero drift approximation,
is self-consistent and closed and entirely determines the
functions $\psi(u)$, $\mupar(u)$ and $\muper(u)$. The correction
of the first drift aproximation to the distribution function does
not affect the magnetic hydrodynamics equation.

\section{Derivation of the magnetic
hydrodynamics equationsè}
\subsection{Algebraic structure of the distribution
function\newline moments}
Returning to the original FR by means of (\ref{1.12}), let us
introduce the following scalars:
\begin{displaymath}
\Par{p} = (p,n)\,; \qquad p^2 =  (v^i v^j - g^{ij}) p_i p_j\,;
\end{displaymath}
\begin{equation}  \label{2.1}
\Per{p} = p^2 - \Par{p}^2 = (v^i v^j - g^{ij} - n^i n^j) p_i
p_j\,,
\end{equation}
where $n^i$ is the unit spacelike vector in the direction of the
magnetic field intensity vector $H_i = v^j
\stackrel{\ast}{F}_{ji}$:
\begin{equation} \label{2.1a}
n_i = \frac{H_i}{H}\,; \qquad H^2 = - (H,H)\,.
\end{equation}
then the distribution function in the zero drift approximation
(\ref{1.22}) takes the form
\begin{equation} \label{2.2}
f_a = \stackrel{0}{f}_a (\mupar^{-2}\Par{p}^2 +
\muper^{-2}\Per{p}^2)\,.
\end{equation}

It is not difficult to show that the moments of this ditribution
are:
\begin{displaymath}
n^i_a(x) = \int\limits_{P(x)}^{} f_a(x,p) p^i dP \,;
\end{displaymath}
\begin{equation}  \label{2.3}
T^{ij}_a(x) = \int\limits_{P(x)}^{} f_a(x,p) p^i p^j dP\,,
\end{equation}
where
\begin{displaymath}
dP = \frac{\sqrt{- g} (2S+1) dp^1 dp^2 dp^3}{(2 \pi \hbar)^3 p_4}
\end{displaymath}
($S$ - is the particle spin), and have the following algebraic
structure:
\begin{equation}  \label{2.4}
n^i_a(u) = n v^i \,;
\end{equation}
\begin{equation}  \label{2.5}
T^{ij}_a (u) = (\varepsilon + \Per{P}) v^iv^j - \Per{P} g^{ij} +
(\Par{P} - \Per{P})n^in^j\,,
\end{equation}
where $n$, $\varepsilon$, $\Par{P}$ and $\Per{P}$ are some scalar
functions of the variable $u$; the indices of the kind of
particles, $a$, are dropped in these scalars for the sake of
simplicity of notations. The EMT track of particles is equal to:
\begin{equation}  \label{2.6}
T_a (u) = \varepsilon - 2 \Per{P} - \Par{P}\,.
\end{equation}
From (\ref{2.4}) it follows that the velocity vector vi introduced
in (\ref{1.11}) coincides with the plasma kinematic velocity
vector. It is not difficult to check the fulfilment of the
condition \cite{gmsw}:
\begin{equation}  \label{2.7}
(v,n) = 0\,,
\end{equation}
and consequently the vector $v^i$ is an eigenvector of the
particle EMT, i.e. it is the dynamic plasma velocity vector
according to Synge \cite{sing}. thus the frame of reference
introduced by the relations (\ref{1.12}) is comoving the plasma.

Note that in the first drift approximation the equation for
kinematic and dynamic velocities of particles is violated and
therefore a drift current arises.

\subsection{Magnetic hydrodynamics equation for an
anisotropic plasma in the PGW field}
In Sec. 3 of Ref. \cite{gmsw} the RMHD equations for a plasma in
an arbitrary gravitational field and for an arbitrary structure of
the plasma EMT, $\stackrel{p}{T}_{ij}$, with the eigenvector $v^i$
were obtained:
\begin{equation}  \label{2.8}
\stackrel{p}{T}_{ij} v^j = \varepsilon v_i \,,
\end{equation}
where the invariant $\varepsilon > 0$ is the plasma energy density
in the comeoving FR. In this case the following conditions were
imposed on the invariants of the electromagnetic field:
\begin{equation}  \label{2.9}
F^{ij} \stackrel{\ast}{F}_{ij} = 0\,;
\end{equation}
\begin{equation}  \label{2.10}
F_{ij} F^{ij} = 2 H^2 > 0\,.
\end{equation}
In \cite{gmsw} it was shown that, under these conditions, from the
conservation law
\begin{equation}  \label{2.11}
T^{ij}_{,j} = 0
\end{equation}
for the whole EMT of the plasma and the electromagnetic field
\begin{displaymath}
T_{ij} = \stackrel{p}{T}_{ij} + \stackrel{f}{T}_{ij}
\end{displaymath}
and the first group of the Maxwell equations
\begin{equation}  \label{2.12}
\stackrel{\ast}{F^{ij}}_{,j} = 0
\end{equation}
it follows:
\begin{enumerate}
\item
The conditions of embedding of the magnetic field in the plasma
are:
\begin{equation}  \label{2.13}
F_{ij} v^j = 0
\end{equation}
(in this case the velocity vector vi also automatically becomes an
eigenvector of the electromagnetic field EMT);
\item
The second group of the Maxwell equations is
\begin{equation}  \label{2.14}
F^{ij}_{,j} = - 4 \pi J^i_{dr}
\end{equation}
with the drift current
\begin{equation} \label{2.15}
J^{i}_{dr} = - \frac{2 F^{ik} \stackrel{p}{T^{l}}_{k,l} }{F_{jm}
F^{jm}} \,;
\end{equation}
\item
The differential relations are
\begin{equation}
\label{2.16} v^{i} \stackrel{p}{T^{k}}_{i,k} =0\,,
\end{equation}
\begin{equation}
\label{2.17} H^{i} \stackrel{p}{T^{k}}_{i,k} =0\,.
\end{equation}
\end{enumerate}

It is not difficult to ascertain that the Maxwell tensor
determined by the vector potential (\ref{1.6}) automatically
satisfies the conditions (\ref{2.10}) and the velocity vector
\ref{1.11}) —-- the embedding conditions (\ref{2.13}). Therefore
in this case, due to the EMT conservation (\ref{2.12}), which is
an exact consequence of the kinetic equations, the relations
(\ref{2.14}) --- (\ref{2.17}) must hold. The isotropic Killing
vector (\ref{1.2}) gives an exact integral of Eqs. (\ref{2.12}):
(\ref{2.12}):
\begin{equation}  \label{2.18}
L^2 T^u_v = \const \,.
\end{equation}
thus, taking into account Eqs. (\ref{1.11}) and (\ref{2.5}) and
the initial conditions (\ref{1.3}) and (\ref{1.7}), we get from
(\ref{2.18}):
\begin{equation}  \label{2.19}
 L^2 (\varepsilon + \Per{P}) = \psi'
(\stackrel{0}{\varepsilon} + \Per{\stackrel{0}{P}}) \Delta (u)\,.
\end{equation}
where we have introduced the so-called {\it GMSW governing
function} (see \cite{gmsw}):
\begin{equation}  \label{2.20}
\Delta (u) = 1 - \alpha^2 ( e^{2 \beta} - 1)
\end{equation}
and the GMSW dimensionless parameter
\begin{equation}  \label{2.21}  \displaystyle
\alpha^2 =  \frac{H^2_0}{4\pi (\stackrel{0}{\varepsilon} +
\Per{\stackrel{0}{P}})} \,.
\end{equation}

It can be further shown that the relation (\ref{2.17}) is
transformed into an identity, and Eq. (\ref{2.16}) gives:
\begin{equation}  \label{2.22}
\varepsilon' + (\ln H)' (\Par{P} - \Per{P}) - \frac{1}{2}
\left(\ln \frac{\psi'}{L^4}\right)' (\varepsilon + \Par{P}) = 0\,.
\end{equation}
The EMT conservation equation and the Maxwell equations do not
give other independent relations.

The missing equations are obtained from the particle number
conservation laws which are also an exact consequence of the
kinetic equations:
\begin{displaymath}
n^i_{,i} = L^{-2} \left( L^2 n v_v \right)' = 0  \Rightarrow
\end{displaymath}
\begin{equation}  \label{2.23}
n_a (u) L^2 = \sqrt{\psi'} \stackrel{0}{n_a} \,.
\end{equation}

Thus we get the set of equations (\ref{2.19}), (\ref{2.22}) and
(\ref{2.23}) for determining the three unknown scalar functions
$\psi$, $\mupar$ and $\muper$. It is only necessary to obtain in
this way explicitly the scalars $n$, $\varepsilon$, $\Par{P}$ and
$\Per{P}$ out of $(1+2n)$ scalars, where $n$ is the particle kind
number.

Note that there are at least two kinds of charged particles (in
the case of interest these are protons and electrons) in the
plasma. Thus, there are two kinds of scalars $\mupar$ and
$\muper$, and one particle number conservation law for a given
kind of particles connects each couple of them (\ref{2.23}). The
summed components of pressure and energy density appear in
equations (\ref{2.19}) and (\ref{2.22}). As a consequence of the
initial electroneutrality of the plasma, (\ref{2.23}), we get a
relation between the local concentrations of neutrons and protons
\cite{probl}:
\begin{equation}  \label{2.23a}
n_e(u) = n_p(u)\,.
\end{equation}

\subsection{Calculating the moments of the
distribution function}
For calculating the above scalar functions it is necessary to find
the moments of the distribution function (\ref{2.3}). An easiest
way to do it is to use Eqs. (\ref{2.4}) and (\ref{2.5}) in the
comoving FR according to (\ref{1.12}), using the property
(\ref{1.13}) of this transformation and using the spherical
coordinates in the momentum space:
\begin{equation}  \label{2.24o}
\begin{array}{lll}
p_1 & = & \muper p \cos \theta \cos \phi \,; \\
p_3 & = & \muper p \cos \theta \sin \phi \,; \\
p_2 & = & \mupar p \sin \theta \,.\\
\end{array}
\end{equation}
In so doing we obtain:
\begin{equation}  \label{2.24}
n = n_0 \muper^2 \mupar \,,
\end{equation}
where
\begin{equation}  \label{2.25}
n_0 = \frac{2S + 1}{2\pi^2} \int\limits_{0}^{\infty}
\stackrel{0}{f}(p^2)p^2 dp \,,
\end{equation}
the integration variable is:
\begin{equation}
p^2 = \mupar^{-2} \Par{p}^2 + \muper^{-2} \Per{p}^2\,.
\end{equation}
thus we obtain from (\ref{2.24}), (\ref{2.23}) and (\ref{2.23a})):
\begin{equation}  \label{2.24a}
(\mupar^2 \muper)_e = (\mupar^2 \muper)_p  =
\frac{\sqrt{\psi'}}{L^2} (\mupar^2 \muper)_0 \,.
\end{equation}

Further, setting for definiteness
\begin{displaymath}
\muper \leq \mupar\,,
\end{displaymath}
we get the expressions for the components of the plasma pressure
and energy density:
\begin{eqnarray} \label{2.26}
\Par{P} = \frac{(2S+1) \muper^2 \mupar^3}{4 \pi^2 (\muminmu)}\int\limits_{0}^{\infty} \stackrel{0}{f} (p^2) p^2 dp\times\hspace{5cm}\nonumber\\
\left[\sqrt{\mcmupar}  - \frac{\mcmuper}{p \sqrt{\muminmu}} \mupln
\right];
\end{eqnarray}
\begin{eqnarray} \label{2.27}
\Per{P} = \frac{(2S+1)\muper^4 \mupar}{8 \pi^2 (\muminmu)}
\int\limits_{0}^{\infty} \stackrel{0}{f} (p^2) p^2
dp\times \left[ - \sqrt{\mcmupar}\right.\hspace{2cm}\nonumber\\
\left. +\frac{m^2 + p^2 (2 \mupar^2 - \muper^2)}{p
\sqrt{\muminmu}} \mupln\right];
\end{eqnarray}
\begin{eqnarray}  \label{2.28}
\varepsilon = \frac{2S+1}{4\pi^2} \muper^2 \mupar
\int\limits_{0}^{\infty} \stackrel{0}{f}(p^2) p^2 dp\times\hspace{5cm} \nonumber\\
\left[ \sqrt{\mcmupar} + \frac{\mcmuper}{p \sqrt{\muminmu}} \mupln
\right] \,.
\end{eqnarray}
Note that all the expressions (\ref{2.26}) -- (\ref{2.28}) have
finite limits at $\muminmu = 0$. From (\ref{2.26}) -- (\ref{2.28})
one can obtain:
\begin{eqnarray}  \label{2.28a}
\varepsilon - \Par{P} - 2 \Per{P} =
\hspace{6cm}\nonumber\\
\frac{(2S+1) m^2 \muper^2 \mupar}{2 \pi^2
\sqrt{\muminmu}}\int\limits_{0}^{\infty}\mupln \stackrel{0}{f}
(p^2) p dp\,.
\end{eqnarray}
Besides, note that in the case $\mupar \leq \muper$ it is
essential that the logarithmic functions in the above formulae for
the moments are to be substituted by the functions $\arcsin$.

\section{Ultrarelativistic plasma}
\subsection{Energy density and pressure}
magnetosphere (see, e.g., \cite{pulsar}), leads to very high
values of the kinetic energy of electrons and protons (up to
$10^{18}$ev). Thus GMSW appear to be always realized in an
ultrarelativistic plasma.

In the ultrarelativistic plasma the particle rest mass does not
affect the macroscopic moments in the drift approximation,
therefore we can set:
\begin{equation}  \label{3.1}
(\mupar)_e = (\mupar)_p \,; \quad (\muper)_e = (\muper)_p \,.
\end{equation}

For an ultrarelativistic plasma,
\begin{equation}  \label{3.2}
\muper p \gg m\,; \quad \mupar p \gg m \,;
\end{equation}
the above expressions become explicit functions of $\muper$ and
$\mupar$:
\begin{equation}  \label{3.3}
\Par{P} = \frac{E_0}{2} \frac{\muper^2 \mupar^3}{\muminmu}\left[
\mupar - \frac{\muper^2}{\sqrt{\muminmu}} \ln \left( \frac{\mupar
+ \sqrt{\muminmu}}{\muper}\right) \right];
\end{equation}
\begin{equation}  \label{3.4}
\Per{P} = \frac{E_0}{4} \frac{\muper^4
\mupar}{\muminmu}\left[\frac{2 \mupar^2 -
\muper^2}{\sqrt{\muminmu}} \ln \left( \frac{\mupar +
\sqrt{\muminmu}}{\muper}\right) - \mupar \right]\,,
\end{equation}
\begin{equation}  \label{3.4}
\varepsilon = \frac{E_0}{2} \muper^2 \mupar\left[ \mupar +
\frac{\muper^2}{\sqrt{\muminmu}} \ln \left( \frac{\mupar +
\sqrt{\muminmu}}{\muper}\right) \right],
\end{equation}
where
\begin{equation}  \label{3.5}
E_0 = \sum\limits_{a}^{} \frac{2S+1}{2 \pi^2}
\int\limits_{0}^{\infty} \stackrel{0}{f} (p^2)p^3 dp  = \const
\end{equation}
is the energy density of an isotropic homogeneous plasma. From
(\ref{2.28a}), in the approximation under study it follows:
\begin{equation} \label{3.6}
\varepsilon - \Par{P} - 2 \Per{P} = 0 \,.
\end{equation}

\subsection{Special solutions}
With the expressions obtained, it is still very difficult to
extract any information from Eqs. (\ref{2.19}) and (\ref{2.22}).
Therefore let us study these equations at the limiting values of
the plasma anisotropy parameter:
\begin{equation}  \label{3.7}
\delta = \frac{\muper}{\mupar}\,.
\end{equation}
\subsubsection{Isotropic plasma: $\delta \rightarrow 1$}
Calcuating the limits of the expressions (\ref{3.2}) - (\ref{3.5})
as $\delta \rightarrow 1$, we get:
\begin{equation}  \label{3.8}
\Par{P} = \Per{P} = \frac{1}{3} \varepsilon = \frac{1}{6} \mu^4
E_0\,,
\end{equation}
where $\mupar = \muper = \mu$. Then it is easy to verify that Eqs.
(\ref{2.19}) and (\ref{2.24a}) have similar solutions:
\begin{equation}  \label{3.9}
\mu^3 = \frac{\mu^3_0}{L^2} \sqrt{\psi'} \,.
\end{equation}
This fact is of fundamental importance as it provides a
microscopic explanation to RMHD \cite{gmsw}. In reality, if Eqs.
(\ref{2.19}) and (\ref{2.24a}) had independent consequences, it
would indicate that the initial anisotropy of a collisionless
plasma is broken by a gravitational wave, i.e. the RMHD equations
would have been ineligible for the description of the
collisionless plasma.

Substituting relation (\ref{3.9}) into Eq. (\ref{2.22}), we get
the known solutions \cite{gmsw} for an isotropic ultrarelativistic
plasma:
\begin{equation}  \label{3.10}
\psi' = \frac{1}{L^2 \Delta^3}\,;
\end{equation}
\begin{equation} \label{3.10a}
\varepsilon = \frac{\varepsilon_0}{L^4 \Delta^2}\,;
\end{equation}
\begin{equation}  \label{3.10b}
H^2 = H^2_0 \frac{e^{2\beta}}{L^4 \Delta^3}\,;
\end{equation}
in this case
\begin{equation}  \label{3.10c}
\mu = \frac{\mu_0}{L} \frac{1}{\sqrt{\Delta}} \,.
\end{equation}

\subsubsection{Plasma with a chilled transverse momentum: $\delta
\rightarrow 0$}
In this case from (\ref{3.2}) - (\ref{3.5}) we obtain:
\begin{equation}  \label{3.11}
\Par{P} \approx \varepsilon = \frac{1}{2} E_0 \mupar^2 \muper^2
\,; \quad \Per{P} \approx 0\,.
\end{equation}
Then Eq. (\ref{2.19}) has an integral:
\begin{equation}  \label{3.12}
\frac{\varepsilon H L^4}{\psi'} = \varepsilon_0 H_0\,.
\end{equation}
Using in (\ref{3.12}) the integral (\ref{2.24a}) and the
expression for $H$ (\ref{1.15}), we obtain an expression for
$\muper$:
\begin{equation}  \label{3.13}
\muper = (\muper)_0 \frac{e^{- \beta}}{L}\,.
\end{equation}
Now, substituting (\ref{3.13}) into (\ref{3.11}) and the result
obtained into (\ref{2.19}), using the integral (\ref{2.24a}), we
get:
\begin{displaymath}
\psi' = \frac{e^{- 2\beta}}{L^2 \Delta^2} \,;
\end{displaymath}
\begin{equation}  \label{3.14}
\mupar = \frac{(\mupar)_0}{L \sqrt{\Delta}}\,;
\end{equation}
\begin{equation}  \label{3.15}
\varepsilon = \Par{P} = \varepsilon_0 \frac{e^{- 2\beta}}{L^4
\Delta}\,; \quad \Per{P} = \Per{\stackrel{0}{P}} \frac{e^{-
4\beta}}{L^4}\,;
\end{equation}
\begin{equation}  \label{3.16}
H^2 = \frac{H^2_0}{L^4 \Delta^2} \,.
\end{equation}
From the solutions presented it is obvious that on the GMSW front
($\Delta \rightarrow 0$) the transverse pressure component is
almost conserved ($\Per{P} \approx \Per{\stackrel{0}{P}})$,
whereas the the longitudinal pressure component tends to infinity
proportionally to $\Delta^{- 1}$. thus, on the GMSW front the
plasma initial anisotropy (in case it exists) intensifies.

\section{Anisotropy effect on the
magneto - \newline bremsstrahlung}
The squared projection values, averaged over the distribution
(\ref{2.2}) in the comoving FR, are of the order
\begin{equation}  \label{4.1}
<\Per{p}^2> \sim \muper^2 \,; \quad <\Par{p}^2> \sim \mupar^2 \,.
\end{equation}
therefore the average value of the ultrarelativistic particle
energy is
\begin{equation}  \label{4.2}
<{\cal E}> = \sqrt{\Per{p}^2 + \Par{p}^2} \sim \sqrt{\muper^2 +
\mupar^2} \,.
\end{equation}
thus in the case of a highly anisotropic plasma (for $\delta
\rightarrow 0$) we obtain from (\ref{4.2}):
\begin{equation}  \label{4.3}
<{\cal E}> \sim \mupar \,.
\end{equation}
Substituting  $\muper$ and $\mupar$ from (\ref{3.13}) and
(\ref{3.14}) into these expressions, we get:
\begin{displaymath}
<\Per{p}^2> \approx <\Per{p}^2>_0 \frac{e^{- 2\beta}}{L^2} \approx
\const \,;
\end{displaymath}
\begin{displaymath}
<\Par{p}^2> \approx \frac{<\Par{p}^2>_0}{L^2 \Delta}\,;
\end{displaymath}
\begin{equation}  \label{4.4}
<{\cal E}> \approx \frac{<{\cal E}>_0}{L \sqrt{\Delta}}\,.
\end{equation}
At small values of $\delta$ the average value of the angle $\chi$
between the vectors of microscopic velocity of an
ultrarelativistic particle ${\bf V}$ and the magnetic field
intensity ${\bf H}$ is small as well:
\begin{equation}  \label{4.5}
\sin \chi = \frac{\Per{V}}{V} \approx
\frac{\sqrt{<\Per{p}^2>}}{<{\cal E}>} \Rightarrow \chi \approx
\chi_0 e^{- \beta} \sqrt{\Delta}\,,
\end{equation}
where $\chi_0 = \delta$.

As is known (see e.g. \cite{land}), the magnetobremsstrahlung
intensity for an ultrarelativistic particle is concentrated in a
narrow cone with the axis ${\bf H}$ and the angle $2 \chi$ at the
top. Thus, if $2\chi_0$ is the output angle of
magnetobremsstrahlung of a nonperturbed plasma, then $2\chi$ is
the output angle of magnetobremsstrahlung in the GMSW. On the GMSW
front ($\Delta(u) \rightarrow 0$) the output angle of the
radiation tends to zero. In this case the total intensity of the
magnetobremsstrahlung of an electron\cite{land}:
\begin{equation}  \label{4.6}
I = \frac{2 e^4}{3 m^4} \Per{p} H^2
\end{equation}
according to (\ref{3.16} and (\ref{4.4}) in the GMSW is equal to:
\begin{equation}  \label{4.7}
I(u) = I_0 \frac{e^{- 2\beta(u)}}{L^6(u) \Delta^2(u)}
\end{equation}
(where $I_0$ is an unperturbed radiation intensity) and tends to
infinity on the GMSW front.

If the output angle of the magnetobremsstrahlung of the pulsar
magnetosphere is connected with the duration of the radiation
pulse, then, as the gravitational wave travels through the pulsar
magnetosphere, a strong contraction of the pulse with a
simultaneous increase of its intensity, will be observed. This
effect offers an Alternative, as compared to \cite{gmsw2} and
\cite{detector}, opportunity to explain the giant pulses of the
pulsars, whose radiation direction diagram is in most cases
pencil-like.

%\small

\end{document}